\documentclass{article}

\usepackage{arxiv}

\usepackage[utf8]{inputenc}
\usepackage[T1]{fontenc}
\usepackage{url}
\usepackage{booktabs}
\usepackage{amsmath}
\usepackage{amsfonts}
\usepackage{nicefrac}
\usepackage{microtype}
\usepackage{graphicx}
\usepackage[svgnames]{xcolor}
\usepackage[colorlinks=true,linkcolor=NavyBlue,citecolor=NavyBlue,urlcolor=NavyBlue]{hyperref}
\usepackage{subcaption}
\usepackage{array}
\usepackage{multirow}
\usepackage{siunitx}
\usepackage{makecell}
\usepackage[font=small]{caption}

\usepackage[authoryear,round]{natbib}
\setcitestyle{aysep={}}

\graphicspath{{./images/}}
\title{SEABAD: A Tropical Bird Activity Detection Dataset for Passive
  Acoustic Monitoring}

\author{
  Muhammad Mun'im Ahmad Zabidi$^{1,2}$ \quad
  Mohd Yamani Idna Idris$^{1}$ \quad
  Norisma Idris$^{1}$ \\[4pt]
  $^{1}$Faculty of Computer Science and Information Technology,
    Universiti Malaya, Kuala Lumpur, Malaysia \\
  $^{2}$Faculty of Electrical Engineering, Universiti Teknologi Malaysia,
    Johor Bahru, Malaysia \\[4pt]
  \texttt{yamani@um.edu.my}
}

\date{}

\begin{document}

\twocolumn[
  \maketitle

  \begin{center}
    \small
    \textbf{Dataset:} \url{https://zenodo.org/records/18290494} \quad
    \textbf{Code:} \url{https://github.com/mun3im/seabad}
  \end{center}

  \begin{abstract}
Passive acoustic monitoring (PAM) enables large-scale biodiversity assessment,
but continuous recording generates large amounts of non-informative audio,
creating challenges for storage, power consumption, and long-term edge deployment.
Bird audio detection (BAD), which identifies bird vocalizations, can reduce this
burden by filtering irrelevant recordings before downstream analysis. However,
most BAD systems are trained on temperate datasets despite tropical soundscapes
being denser, more species-rich, and acoustically unpredictable.

To address this gap, we introduce SEABAD (Southeast Asian Bird Activity Detection),
a dataset of 50,000 curated three-second clips from Southeast Asian soundscapes,
evenly balanced between bird-present and bird-absent samples. The dataset spans
1,677 bird species and is standardized to 16~kHz mono audio for embedded and
low-power inference. We developed a dual-branch curation pipeline: a six-stage
positive-label workflow applied to Xeno-Canto recordings, alongside six
source-specific negative-label extractions from environmental datasets. These
procedures reduced class imbalance by 13.7\% (Gini coefficient: $0.601 \rightarrow 0.519$).
A manual audit of 1,000 positive clips confirmed $97.8\% \pm 0.9\%$ labeling accuracy.
Baseline experiments using MobileNetV3-Small achieved $99.57\% \pm 0.25\%$ accuracy
and $0.9985 \pm 0.0002$ AUC across three random seeds. SEABAD and the full curation
pipeline are publicly released to support tropical BAD research and energy-efficient
acoustic monitoring.
  \end{abstract}


  \bigskip
]


\section{Introduction}

Tropical ecosystems support more than half of the world's terrestrial
biodiversity, yet they are also among the regions undergoing the fastest
ecological change and habitat degradation. Because birds respond sensitively
to shifts in vegetation structure, habitat quality, and ecosystem stability,
their vocal activity is widely used as an indicator of broader environmental
conditions (\citealt{fraixedas2020state}; \citealt{sethi2022soundscapes}).
This has made long-term avian monitoring increasingly important for
conservation planning, ecological assessment, and biodiversity research.

Traditional survey methods such as point counts and transect observations
remain valuable, but they are difficult to scale and depend heavily on
observer availability and expertise (\citealt{chauvier2021novel};
\citealt{perez2021estimating}). These challenges become even more pronounced
in tropical forests, where dense canopy cover, uneven terrain, and
exceptionally high species diversity complicate direct observation
(\citealt{mendoza2023past}; \citealt{crunchant2021acoustic};
\citealt{gibbons2026monitoring}). In response, passive acoustic monitoring
(PAM) has emerged as a widely adopted alternative. By continuously recording
environmental soundscapes over long periods, PAM enables non-invasive
monitoring across locations and timescales that would be difficult to survey
manually (\citealt{priyadarshani2018automated}; \citealt{melo2021active};
\citealt{ross2023passive}).

The practical problem is that most of these recordings contain very little
useful information. Autonomous recording units (ARUs) are commonly configured
to record at fixed intervals or respond to simple amplitude thresholds, which
means large portions of the collected audio consist of wind, rainfall,
insects, or other background noise rather than bird vocalizations. In some
deployments, bird activity accounts for less than 10\% of the recorded
material (\citealt{huang2024tinychirp}). For field systems operating under
tight storage and energy budgets, this inefficiency reduces deployment
duration and substantially increases the effort required for downstream
processing and annotation.

Recent progress in TinyML has made it possible to run neural inference
directly on low-power microcontroller platforms (\citealt{banbury2021micronets};
\citealt{david2021tensorflow}). This creates an opportunity to move beyond
continuous indiscriminate recording toward more selective acoustic sensing.
Within this setting, Bird Activity Detection (BAD) serves as a lightweight
front-end task that determines whether bird vocalizations are present in an
audio segment. Non-informative recordings can then be discarded before storage
or transmission. Compared with full species classification, binary
presence--absence detection can be implemented using relatively compact models
that fit within the memory and energy constraints of devices such as AudioMoth
recorders, making BAD a practical strategy for improving PAM efficiency
(\citealt{stowell2014open}; \citealt{stowell2018automatic};
\citealt{da2019evaluation}; \citealt{lostanlen2019robust}).

However, progress in low-power BAD is constrained less by model architecture
than by data availability and dataset construction practices. Most publicly
available bird audio datasets originate from temperate regions, and models
trained on these corpora often generalize poorly to tropical soundscapes
(\citealt{burivalova2022loss}). Tropical environments present distinct
acoustic conditions, including high species richness, frequent multi-species
choruses, and dense non-avian biophony from sources such as cicadas and
primates. Together, these factors alter soundscape statistics and reduce
detection performance when models are transferred from temperate to tropical
domains. Existing datasets are also rarely designed for embedded
presence--absence detection, and few studies provide reproducible pipelines
for building region-specific BAD datasets.

Current datasets therefore do not fully satisfy the requirements for
edge-deployable BAD in Southeast Asia. In particular, few offer short
fixed-duration clips compatible with microcontroller buffer constraints
(3\,s at 16\,kHz), binary presence--absence labels rather than species-level
annotations, and broad representation of Southeast Asian tropical soundscapes.
These requirements are seldom addressed together within a single dataset,
creating a mismatch between embedded inference constraints and the acoustic
complexity of tropical ecosystems.

To address this gap, we introduce \textbf{SEABAD (Southeast Asian Bird
Activity Detection)}, a large-scale tropical dataset and a reproducible
framework for constructing bird presence--absence corpora. SEABAD consists
of 50{,}000 automatically curated 3-second clips spanning more than 1{,}600
species across Southeast Asian ecosystems, standardized to 16\,kHz mono for
embedded deployment. Alongside the dataset, we present an end-to-end
curation pipeline covering data acquisition, automated segment extraction,
acoustic deduplication, and diversity-aware species balancing.

\paragraph{Contributions.}
\begin{enumerate}
  \item \textbf{A tropical BAD dataset}: SEABAD, a 50{,}000-clip dataset
        spanning 1{,}677 Southeast Asian species, explicitly designed for
        edge-deployable bird presence detection.
  \item \textbf{A reproducible dual-branch pipeline}: A six-stage
        positive-label workflow (metadata acquisition through quality
        assurance) running in parallel with six source-specific
        negative-label extractions, jointly producing 50{,}000 clips.
  \item \textbf{A diversity-aware balancing method}: A strategy that reduces
        long-tailed species imbalance by 13.7\% (Gini coefficient) while
        preserving intra-species acoustic variability.
  \item \textbf{Strong baseline performance}: Lightweight CNNs achieve
        $>$99\% AUC, indicating high label quality and task separability.
\end{enumerate}

\section{Related work}

\subsection{Hardware and sensing infrastructure}

While proprietary systems like Wildlife Acoustics Song Meter dominate
conservation deployments (\citealt{perez2019cost}; \citealt{mennill2024field}),
their cost limits accessibility. Open-source alternatives like AudioMoth
(\citealt{hill2019audiomoth}; \citealt{bota2023hearing}) combine low-cost
design with energy-efficient recording using Goertzel filters for
frequency-band monitoring. Recent advances enable on-board neural inference
on AudioMoth's Cortex-M4 processor (\citealt{ciapponi2025enabling}), marking
a shift toward edge-based acoustic intelligence.

\subsection{Bird activity detection: algorithms and datasets}

Bioacoustic analysis has shifted from feature-based methods (MFCCs with SVMs;
\citealt{mutanu2022review}) to deep CNNs that learn spectro-temporal
representations (\citealt{stowell2022computational}). Most research focuses
on species classification (e.g., BirdNET; \citealt{kahl2021birdnet}), while
BAD or binary presence detection remains underexplored despite its importance
for edge preprocessing.

Early BAD methods used spectrogram segmentation (\citealt{potamitis2014automatic})
and handcrafted acoustic features (\citealt{zhao2017automated}), requiring
manual feature engineering. The DCASE 2018 challenge
(\citealt{stowell2018automatic}) standardized BAD benchmarking with
Freefield1010, Warblr, and BirdVox-DCASE datasets (10-second clips), but
introduced weak labeling (bird presence $<$5\% of clip) and temporal mismatch
with modern 3-second architectures, complicating edge deployment
(\citealt{solomes2020efficient}). Follow-on work demonstrated that training
across mixed-source datasets improves generalization to unseen recording
conditions (\citealt{kim2020animal}), motivating the heterogeneous negative
sample pool used in SEABAD.

Recent studies have explored finer-grained localization using temporal
segmentation (\citealt{cohen2022tweetynet}) and patch-based CNN architectures
(\citealt{lostanlen2024birdvoxdetect}). Although effective, these methods
depend on dense convolutional processing, making them difficult to run on
highly constrained hardware such as the AudioMoth platform, which operates
within 256\,kB of RAM and strict real-time latency limits of roughly 100\,ms.
Frameworks designed specifically for embedded BAD, such as TinyChirp
(\citealt{huang2024tinychirp}), show that careful optimization can make edge
deployment feasible, but the trade-off between computational cost and
detection reliability remains an open challenge.

\subsection{Geographic bias in bioacoustic datasets}

Global biodiversity data remain heavily imbalanced. More than 70\% of
occurrence records come from Europe and North America, even though these
regions account for less than one-third of the Earth's land surface
(\citealt{hughes2021sampling}). The same pattern appears in bioacoustic
datasets, where recording efforts and sensor deployments are concentrated in
temperate regions. As a result, models trained on these datasets often perform
poorly in tropical soundscapes (\citealt{stowell2019automatic};
\citealt{winiarska2024detection}).

Tropical environments differ substantially in their acoustic structure, with
higher species richness, overlapping multi-species choruses, and dense
non-avian biophony from insects and primates. These differences make it
difficult for models trained in temperate regions to generalize effectively,
highlighting the need for region-specific datasets (\citealt{jeliazkov2022sampling};
\citealt{meyer2016multidimensional}).

\subsection{Dataset annotation strategies}

One of the main differences between bird audio datasets is how the recordings
are annotated. Broadly, existing datasets tend to follow three approaches.
Some rely on \textbf{crowdsourced soft labels}, where contributors indicate
only whether a species is present or absent, as seen in Freefield1010 and
Warblr. Others use \textbf{expert hard labels}, where trained annotators mark
precise temporal boundaries for vocalizations, as in BirdVox-DCASE. A third
category combines automatic labeling with partial manual verification, as in
BirdSet. Table~\ref{tab:dataset_comparison_literature} compares SEABAD with
representative datasets.

SEABAD follows a hybrid strategy. Positive samples inherit species labels from
crowd-sourced metadata available in Xeno-Canto, after which recordings are
segmented automatically using RMS-based energy detection to extract fixed
3-second regions centered on periods of strong vocal activity. This design
retains the scalability of community-generated archives while introducing
temporal consistency suitable for detection tasks.

Recordings with Xeno-Canto quality ratings of A or B are strongly prioritized
during selection --- 92.1\% of the final dataset carries these ratings --- and
the segmentation process preferentially selects regions with the highest
acoustic energy, which are more likely to contain the intended vocalization.

\begin{table*}[t]
  \centering\small
  \caption{Comparison of bird audio datasets for detection and classification
    tasks. SEABAD uses crowd-sourced soft-labeling and automatic
    hard-labeling/segmentation.}
  \label{tab:dataset_comparison_literature}
  \begin{tabular}{@{} l l l r r r l @{}}
    \toprule
    \textbf{Dataset} & \textbf{Region} & \textbf{Task}
      & \textbf{Clips} & \textbf{Species}
      & \textbf{Clip length} & \textbf{Annotation} \\
    \midrule
    Freefield1010 (\citealt{stowell2014open})
      & Global    & Detection      & 7{,}690    & unlabeled & 10\,s    & Crowd  \\
    Warblr (\citealt{stowell2018automatic})
      & UK        & Detection      & 10{,}000   & 20+       & 10\,s    & Crowd  \\
    BirdVox-DCASE (\citealt{lostanlen2018birdvox})
      & N.\ America & Detection   & 20{,}000   & $<$10     & 10\,s    & Expert \\
    BirdSet (\citealt{rauch2025birdset})
      & Global    & Classification & 520{,}000+ & 10{,}000+ & variable & Mixed  \\
    LifeCLEF (\citealt{joly2018overview})
      & Global    & Classification & large      & 1{,}000+  & variable & Mixed  \\
    \midrule
    \textbf{SEABAD}
      & \textbf{SE Asia} & \textbf{Detection}
      & \textbf{50{,}000} & \textbf{1{,}600+}
      & \textbf{3\,s} & \textbf{Mixed} \\
    \bottomrule
  \end{tabular}
\end{table*}

\section{Materials and methods}

\begin{figure}[t]
  \centering
  \includegraphics[scale=.9]{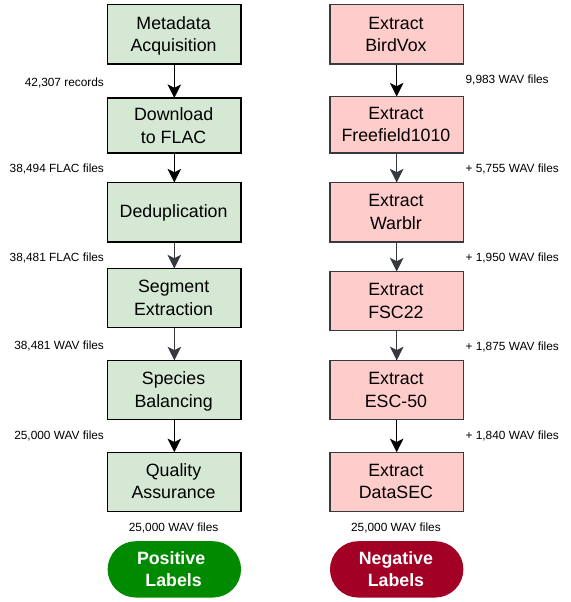}
  \caption{Dual-branch curation pipeline used to construct SEABAD.
    The positive-label branch (left) processes Xeno-Canto recordings from five
    Southeast Asian countries through six sequential stages: metadata acquisition,
    download to FLAC, acoustic deduplication, segment extraction, diversity-aware
    species balancing, and quality assurance, yielding 25{,}000 bird-present clips.
    The negative-label branch (right) runs six parallel source-specific
    extractions from BirdVox-DCASE-20k, Freefield1010, Warblrb10k, FSC-22,
    ESC-50, and DataSEC, yielding 25{,}000 bird-absent clips. Both branches
    produce 3-second, 16\,kHz mono WAV files; the combined 50{,}000-clip dataset
    is used for downstream validation (not shown).}
  \label{fig:pipeline}
\end{figure}

The dataset curation procedure (Figure~\ref{fig:pipeline}) consists of
preparing positive (bird present) and negative (bird absent) labels. For
positive labels, sounds are sourced from the Xeno-Canto open-access
repository and processed using the procedure described below. For negative
labels, non-bird sounds are sourced from the BirdCLEF dataset (Freefield1010,
Warblr, and BirdVox), ESC-50 (Environmental Sound Classification), and FSC-22
(Forest Sound Classification).

\subsection{Positive label curation}

Curating high-quality, ecologically diverse training data from community
repositories such as Xeno-Canto poses several challenges. Raw downloads
exhibit (1) acoustic duplicates due to re-uploads and cross-border species,
(2) severe class imbalance with a long-tailed species distribution, and
(3) limited intra-species diversity caused by repeated use of the same
recordings.

To address these issues, we developed a six-stage positive-label curation
pipeline featuring novel algorithms for acoustic deduplication and
diversity-aware species balancing. The positive-label branch processes
Xeno-Canto recordings through the following stages:

\begin{enumerate}
  \setlength{\itemsep}{0pt}
  \setlength{\parskip}{0pt}
  \item metadata acquisition,
  \item download and conversion to FLAC,
  \item acoustic deduplication,
  \item segment extraction,
  \item species balancing, and
  \item quality assurance.
\end{enumerate}

\subsubsection{Metadata acquisition}

Species metadata and recording information were retrieved using the Xeno-Canto
API (v3), which provides a unique identifier, species taxonomy, recording
quality rating, geographic coordinates, and download links for each recording.
All taxa cataloged as birds were included, with coverage determined entirely
by recording availability on Xeno-Canto.

SEABAD focuses on the Sundaland biogeographic region and southern Indochina,
comprising Malaysia, Indonesia, Singapore, Brunei, and Thailand. These
countries share substantial avifaunal overlap resulting from historical
Pleistocene land-bridge connectivity and collectively represent approximately
1{,}677 bird species. This regional focus captures a large portion of
Southeast Asia's tropical lowland avifauna while maintaining ecological and
biogeographic coherence. Recordings from additional Southeast Asian countries
(e.g., the Philippines, Vietnam, Laos, Cambodia, and Myanmar) were available
but excluded from this release to maintain geographic consistency. The
curation pipeline is designed to support future expansion to a broader
pan-Southeast Asian dataset.

The API query returned 43{,}108 records labelled as \texttt{grp = "birds"}.
After removing 801 entries with unresolved species identities (e.g.,
``unknown'' or ``identity unknown''), 42{,}307 records with valid species
labels were retained.

\subsubsection{Sound acquisition and resampling}

Only entries with valid and accessible audio URLs were downloaded. Recordings
that were blocked, unavailable, or shorter than 3 seconds were excluded. All
audio files (downloaded from Xeno-Canto in MP3 format) were converted to
mono 16\,kHz FLAC using \texttt{ffmpeg} with polyphase anti-aliasing
resampling for lossless intermediate processing. Preprocessing was
deliberately minimal --- limited to resampling and format conversion --- to
emulate real edge recording conditions. No amplitude normalization or signal
conditioning was applied.

This procedure yielded 38{,}494 usable recordings. A total of 3{,}799 entries
were excluded due to missing URLs or insufficient duration, and 14 downloads
failed to convert after download.

\subsubsection{Acoustic similarity-based deduplication}

Community-contributed repositories can contain duplicate recordings due to
re-uploads, metadata revisions, or format conversions. Such duplicates pose a
risk of train-test contamination when they span data splits. Traditional
hash-based methods fail to detect acoustically identical files stored with
different codecs or bitrates. We therefore implemented acoustic
similarity-based deduplication using mel-spectrogram embeddings and FAISS
(\citealt{johnson2019billion}) approximate nearest-neighbor search.

Each 3-second recording was converted to a 128-bin mel spectrogram (16\,kHz,
512-point FFT, 128-sample hop). A compact 256-dimensional embedding
$\mathbf{z} = [\mu(\mathbf{E}), \sigma(\mathbf{E})]$ was computed by
concatenating mean and standard deviation across time frames, then
L2-normalized for cosine similarity comparison. FAISS indexed all candidate
embeddings, retrieving the top-6 nearest neighbors per clip, avoiding
$O(N^2)$ exhaustive comparison.

Duplicate pairs were identified when embeddings were identical within machine
precision ($\|\mathbf{z}_i - \mathbf{z}_j\|_2 < 10^{-7}$), which occurs when
two clips produce numerically indistinguishable mel-spectrogram statistics.
This conservative criterion avoids false duplicate removal while still
detecting exact acoustic duplicates. For each duplicate pair, the file with
the higher Xeno-Canto catalog number was removed, preserving earlier uploads.

\subsubsection{Segment extraction}

Recordings were segmented into fixed-length 3\,s clips to capture complete
vocalizations while maintaining computational efficiency. Each recording was
analyzed using a sliding window (3\,s window, 100\,ms step) with RMS
amplitude computed for each position:

\[
\mathrm{RMS} =
\sqrt{\frac{1}{N}\sum_{n=0}^{N-1} y[n]^2},
\]

where $y[n]$ denotes waveform samples and $N$ the window length. Windows
with $\mathrm{RMS} \geq 0.001$ were ranked by energy, and clips selected
while enforcing 1.5\,s minimum temporal separation to prevent redundancy.
For recordings $>$12\,s, the first 3\,s were skipped to avoid voice
annotations or handling noise. Clips with amplitude clipping
(peak $\geq$ 0.9999) were corrected using peak scaling and soft limiting.
Full-retention mode preserved all extracted clips for subsequent
species-level balancing.

\subsubsection{Species balancing}
\label{sec:balanced_dataset}

\paragraph{Problem formulation}

Community-contributed archives exhibit long-tail species distributions where
common species dominate (\citealt{stowell2018automatic};
\citealt{kahl2021overview}). This biases models toward overrepresented
species, reducing performance on rare taxa critical for biodiversity
monitoring (\citealt{he2009learning}; \citealt{priyadarshani2018automated};
\citealt{benkendorf2023correcting}).

After deduplication, 38{,}481 clips spanning 1{,}677 species showed high
inequality (mean 22.9 clips/species, Gini 0.601). Balancing required reducing
this inequality while preserving acoustic diversity within species and
recording-level diversity across sources.

\paragraph{Related balancing strategies}

Several strategies have been proposed to address imbalance in bioacoustic
datasets. In general machine learning, random undersampling and synthetic
oversampling techniques such as SMOTE are commonly used (\citealt{he2009learning};
\citealt{chawla2002smote}). However, SMOTE-style synthesis is unsuitable for
ecological datasets because artificial vocalizations would not correspond to
real biological recordings.

Within bioacoustics, balancing is typically achieved through simple per-species
caps, as widely practiced in BirdCLEF challenge pipelines (\citealt{joly2018overview};
\citealt{kahl2021overview}). BirdVox-70k (\citealt{lostanlen2018birdvox})
limits clip density per recording to prevent near-duplicate segments.
BirdSet (\citealt{rauch2025birdset}) applies quality filtering and multi-stage
curation analogous to our approach. NIPS4Bplus (\citealt{morfi2018nips4bplus})
accepts class imbalance to reflect natural soundscapes. Our approach extends
these practices by combining recording-level diversity, large-scale Xeno-Canto
curation, and ecological long-tail awareness with explicit acoustic clustering
to ensure selected samples represent diverse call types within each species.

\paragraph{Acoustic diversity-aware undersampling}

We implemented an acoustic diversity-aware stratified undersampling procedure
that combines species-level balancing with within-species acoustic clustering
and quality-aware selection. The algorithm operates in three stages.

First, an \textit{acoustic salience score} is computed for each clip:

\begin{equation}
  \text{salience} = 0.7 \cdot \frac{\text{mean\_contrast}}{40.0} +
  0.3 \cdot \frac{\text{mean\_centroid}}{f_s}
\end{equation}

This score prioritizes clips with clear foreground vocalizations and
penalizes low-energy or noisy segments.

Second, clips for each species are grouped using MiniBatch K-Means clustering
applied to mel-spectrogram embeddings into five acoustic clusters representing
distinct vocal behaviors (e.g., songs, calls, or different call types). Given
a target dataset size $N_\text{target}$ and $S$ species, the base per-species
allocation is:

\begin{equation}
  n_{\text{base}} = \left\lfloor \frac{N_\textrm{target}}{S} \right\rfloor
\end{equation}

For species with fewer than $n_{\text{base}}$ clips, all samples are retained.
For species exceeding this threshold, representative clips are selected across
acoustic clusters, prioritizing those with higher salience.

Third, a priority-queue backfilling stage expands the dataset to the target
size by selecting additional clips ranked by:

\begin{equation}
  \text{score} =
  \text{salience} +
  \text{quality\_bonus} +
  \text{diversity\_bonus}
\end{equation}

where quality bonuses favor A/B-rated recordings and diversity bonuses
prioritize clips from previously unused acoustic clusters. If the resulting
set exceeds $N_\text{target}$, a final global trimming step retains the
highest-salience clips.

\paragraph{Gini coefficient}

To quantify dataset inequality, we compute the Gini coefficient
(\citealt{gini1909concentration}; \citealt{giorgi2016gini}):

\begin{equation}
  G =
  \frac{2 \sum_{i=1}^{S} i n_i}{S \sum_{i=1}^{S} n_i}
  -
  \frac{S+1}{S}
\end{equation}

where $n_i$ denotes the sample count of the $i$-th species sorted in
ascending order. Values range from 0 (perfect equality) to 1 (maximum
inequality). This metric is widely used to measure ecological evenness and
class imbalance in biodiversity datasets (\citealt{wittebolle2009initial};
\citealt{daly2018ecological}).

\subsection{Negative label curation}

Robust BAD requires carefully curated negative examples that capture the
acoustic complexity of natural environments. False positives from weather
events, insect choruses, machinery, and human activity can severely degrade
field performance, particularly in tropical deployments where soundscapes
exhibit high temporal and spectral variability. We assembled 25{,}000
bird-absent clips from multiple open-access datasets
(Table~\ref{tab:negative_sources}), ensuring both acoustic diversity and
ecological realism.

\begin{table*}[t]
  \centering\footnotesize
  \caption{Public sound datasets used to extract non-bird samples.}
  \label{tab:negative_sources}
  \setlength{\tabcolsep}{3.5pt}
  \begin{tabular}{@{} l c c c c c c l @{}}
    \toprule
    \textbf{Dataset} & \textbf{Duration}
      & \textbf{\makecell{Bird\\classes}}
      & \textbf{\makecell{Non-bird\\classes}}
      & \textbf{\makecell{Bird\\samples}}
      & \textbf{\makecell{Non-bird\\samples}}
      & \textbf{\makecell{Usable\\samples}}
      & \textbf{Geography} \\
    \midrule
    BirdVox-DCASE-20k  & 10\,s  & 1 & 1  & 10{,}017 & 9{,}983 & 9{,}983 & NE USA           \\
    Freefield1010       & 10\,s  & 1 & 1  & 1{,}935  & 5{,}755 & 5{,}755 & Global           \\
    Warblr              & 10\,s  & 1 & 1  & 6{,}045  & 1{,}955 & 1{,}950 & UK               \\
    FSC-22              & 5\,s   & 3 & 27 & 150      & 1{,}875 & 1{,}875 & European forests \\
    ESC-50              & 5\,s   & 4 & 46 & 160      & 1{,}840 & 1{,}840 & Global           \\
    DataSEC             & varies & 3 & 19 & 233      & 4{,}059 & 3{,}597 & Southern Europe  \\
    \midrule
    \multicolumn{6}{r}{\textbf{Total negative samples}} & \textbf{25{,}000} & \\
    \bottomrule
  \end{tabular}
\end{table*}

\subsubsection{Curation strategy and preprocessing}

\paragraph{DCASE bird detection datasets}

The foundation of our negative samples derives from DCASE 2018 bird detection
benchmarks: BirdVox-DCASE-20k (\citealt{lostanlen2018birdvox}), Freefield1010
(\citealt{stowell2014open}), and Warblrb10k (\citealt{stowell2018automatic}).
These datasets provide expert-verified \textit{hasbird}$=$0 annotations from
realistic outdoor conditions --- wind, rain, insects, distant anthropogenic
sounds, and ambient noise --- precisely the acoustic events that commonly
trigger false alarms in automated detection systems
(\citealt{stowell2018automatic}; \citealt{dcase2018task}).

The DCASE 2018 task comprises three development sets (Freefield1010 with
7{,}690 clips, Warblrb10k with 8{,}000 clips, and BirdVox-DCASE-20k with
20{,}000 clips) plus three evaluation sets. From these, we extracted 17{,}685
verified negative examples. DCASE's 10-second clip format introduces a
temporal mismatch with modern 3-second architectures; we reprocessed all
recordings into centered 3-second segments to harmonize temporal resolution
and align with edge-AI deployment requirements.

\paragraph{FSC-22 forest sound classification}

The FSC-22 dataset (\citealt{bandara2023forest}) was used to add ecologically
realistic non-avian biological sounds from both tropical and temperate forest
environments. After removing avian categories (\texttt{BirdChirping},
\texttt{WingFlapping}), 1{,}875 samples containing mammals, amphibians,
insects, rainfall, and wind were retained. These sounds are commonly present
in forest monitoring recordings but are often underrepresented in bird
detection benchmarks.

\paragraph{ESC-50 environmental sounds}

ESC-50 (\citealt{piczak2015esc}; \citealt{fonseca2017freesound}) provides
2{,}000 environmental recordings across 50 semantic classes, contributing
urban, mechanical, and human-made distractors such as footsteps, engines,
doors, and machinery. After excluding avian classes (\texttt{chirping\_birds},
\texttt{crow}, \texttt{rooster}, \texttt{hen}), 1{,}840 samples remained.

\paragraph{DataSEC Mediterranean soundscapes}

The DataSEC dataset (\citealt{bandara2023forest}) provides recordings from
Mediterranean agricultural and rural environments in southern Europe. After
excluding avian-related categories (\texttt{Birds}, \texttt{Chicken coop},
\texttt{Crows seagulls and magpies}), the remaining recordings include
vehicles, aircraft, machinery, sirens, thunder, bells, insects, cats, dogs,
and human speech. Recordings shorter than 3\,s were zero-padded to the target duration rather
than discarded; retaining these short clips provided sufficient samples to
meet the per-source quota without drawing from the music category, which was
excluded from the corpus. Longer recordings were segmented by selecting the
3-second window with highest RMS energy. This produced 3{,}597 Mediterranean
soundscape clips.

\subsubsection{Preprocessing pipeline}

All negative samples underwent standardized preprocessing:

\begin{enumerate}
  \item \textbf{Resampling:} All recordings resampled to 16\,kHz mono.
  \item \textbf{Segmentation:} Clips trimmed or centered to exactly 3 seconds.
  \item \textbf{Dynamic range:} No compression applied; clipping artifacts
        were minimal across datasets.
  \item \textbf{Metadata logging:} Each clip cataloged with source file,
        dataset origin, start timestamp, and RMS amplitude in a unified CSV.
\end{enumerate}

\subsubsection{Quality filtering}

\textbf{Phase 1: Quality filtering.} We applied strict quality criteria to
3-second center clips:
\begin{align}
  \nonumber \text{RMS} &\geq 0.0001 \\
  \nonumber \text{Peak amplitude} &\leq 0.98 \\
  \nonumber \text{Dynamic range} &\geq 0.1
\end{align}

These thresholds remove silent clips, clipped or distorted audio, and signals
with insufficient variation.

\textbf{Phase 2: Deployment-aware allocation.} For multi-category datasets
(ESC-50, FSC-22, DataSEC), avian classes were first removed to preserve
negative-sample purity. Remaining categories were prioritised by ecological
relevance: environmental and outdoor recordings were selected first, followed
by voice and human-activity sounds. Recordings shorter than 3\,s were
zero-padded rather than discarded, as this increased the yield of
ecologically realistic samples and avoided the need to draw from the music
category, which was excluded from the final corpus as its structured
harmonic content is rarely encountered in field PAM deployments.

\textbf{Phase 3: Diversity maximization.} For retained categories with large
clip counts (e.g., voices in DataSEC), a random subset was drawn without
replacement to cap per-category contribution and prevent any single sound
type from biasing the trained model.

\subsection{Quality assurance}

\subsubsection{Audit procedure}

\begin{figure*}[t]
  \centering
  \includegraphics[width=0.95\textwidth]{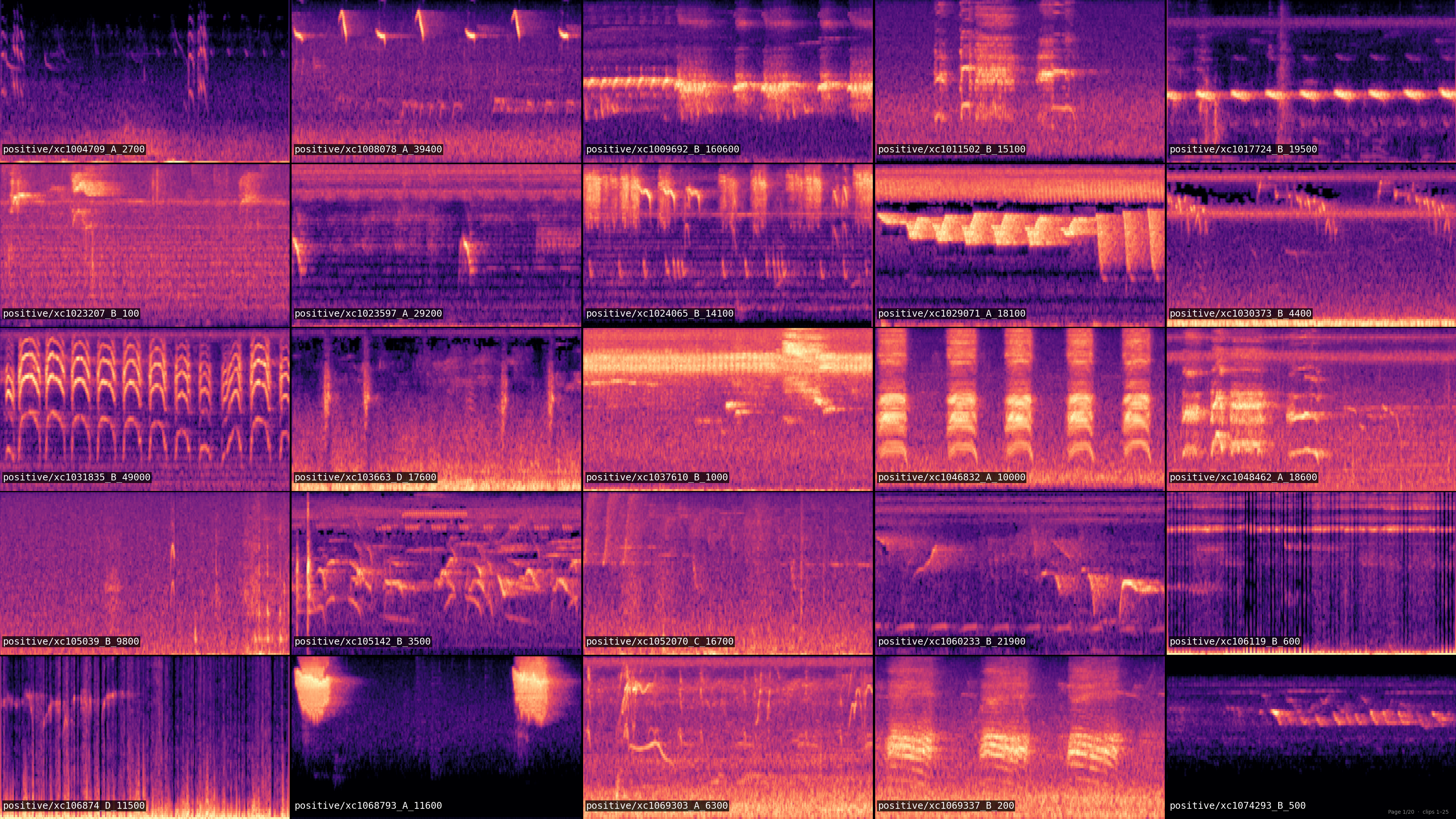}
  \caption{Quality assurance spectrograms showing acoustic diversity
    across SEABAD positive samples. This 5$\times$5 grid was generated at
    4K resolution (3840$\times$2160) for manual auditing on high-resolution
    displays. Examples show varied bird vocalizations including tonal calls,
    trills, repeated note sequences, and complex song structures across Southeast
    Asian species. Spectrograms computed with 80 mel bins, 512 FFT size, 128 hop
    length, spanning 0--8\,kHz over 3-second clips sampled at 16\,kHz.}
  \label{fig:qa_spectrograms}
\end{figure*}

To assess the labeling accuracy of the automated curation pipeline, two
independent random samples totaling 1{,}000 positive clips were inspected by
the first author using mel-spectrogram visualizations and audio playback
(Figure~\ref{fig:qa_spectrograms}). Each audit round presented 500 clips as
20 pages of 5$\times$5 spectrogram grids (25 clips per page) rendered at
4K resolution, enabling efficient batch inspection while preserving visual
detail. Each clip was assigned one of four outcomes: \textit{correct}
(salient bird vocalization visible and audible), \textit{wrong onset}
(vocalization present in the source recording but outside the extracted
3\,s window), \textit{noise dominated} (broadband environmental noise masks
the vocalization), or \textit{no bird} (no vocalization detectable).

Clips affected by onset errors were remediated using an interactive correction
tool that displays the source FLAC spectrogram and waveform, allowing manual
adjustment of the 3\,s extraction window. Corrected onset times were recorded
in a structured log, then re-extracted programmatically to ensure
reproducibility. Clips containing no detectable vocalization ($n=1$) were
removed and replaced by re-running segment extraction on unused regions of the
same source recordings.

Negative clips inherit their labels from the original source datasets (DCASE,
ESC-50, FSC-22, DataSEC), which were independently verified by their
respective authors. No additional manual audit of negative clips was performed
beyond the RMS and peak-amplitude quality filters.

\subsubsection{Audit sample size}

The required audit sample size was determined using Cochran's formula for
estimating proportions in finite populations (\citealt{cochran1977sampling}).
Using the observed error rate from the first audit round ($\hat{p} = 0.04$),
a target margin of error $e^* = 0.015$ (±1.5 percentage points), and 95\%
confidence ($z = 1.96$), the required sample size is:

\begin{equation}
  n_0 = \frac{z^2 \hat{p}(1-\hat{p})}{e^{*2}} = 656, \quad
  n^* = \frac{n_0}{1 + n_0/N} = 639,
\end{equation}

where $N = 25{,}000$ is the population size. A second audit round of 500
clips (total $n = 1{,}000 > 639$) was therefore conducted. This sampling
strategy follows established practice in machine learning dataset quality
auditing (\citealt{northcutt2021pervasive}).

\subsection{Baseline validation experimental setup}

To validate dataset quality, four representative CNN architectures were
evaluated on the binary bird presence--absence detection task. The 50{,}000-clip
dataset was partitioned into training (40{,}000 clips, 80\%), validation
(5{,}000 clips, 10\%), and test (5{,}000 clips, 10\%) sets with stratified
sampling to ensure balanced class distribution across all splits (50\%
positive, 50\% negative).

All models employ transfer learning from ImageNet-pretrained weights, adapting
the final classification layer for binary bird presence detection. Input
mel-spectrograms are computed at 224$\times$224 resolution with 224 mel bins,
1024-point FFT, and hop length calculated to achieve exactly 224 time frames
over 3-second clips sampled at 16\,kHz. While this high-resolution
representation exceeds typical embedded deployment constraints, it enables
thorough validation using standard computer vision architectures.

Training employs Adam optimizer with cosine learning rate decay (initial rate
$1 \times 10^{-4}$), batch size 32, and early stopping with patience 15 epochs
on validation loss. All models apply ImageNet preprocessing: grayscale
mel-spectrograms are replicated to RGB channels, scaled to [0, 255], then
processed with architecture-specific normalization. Models were trained on an
NVIDIA GTX 1080 Ti GPU using TensorFlow 2.15. All audio was resampled to
16\,kHz mono using librosa (v0.10.1) with a Kaiser-windowed sinc resampler.
All reported metrics were computed on the held-out test set and averaged across
three random seeds (42, 100, 786) to verify stability. Complete experimental
code is available at \url{https://github.com/mun3im/mybad/tree/main/validation}.

\section{Results}

\subsection{Dataset composition}

\begin{figure*}[t]
  \centering
  \includegraphics[width=\linewidth]{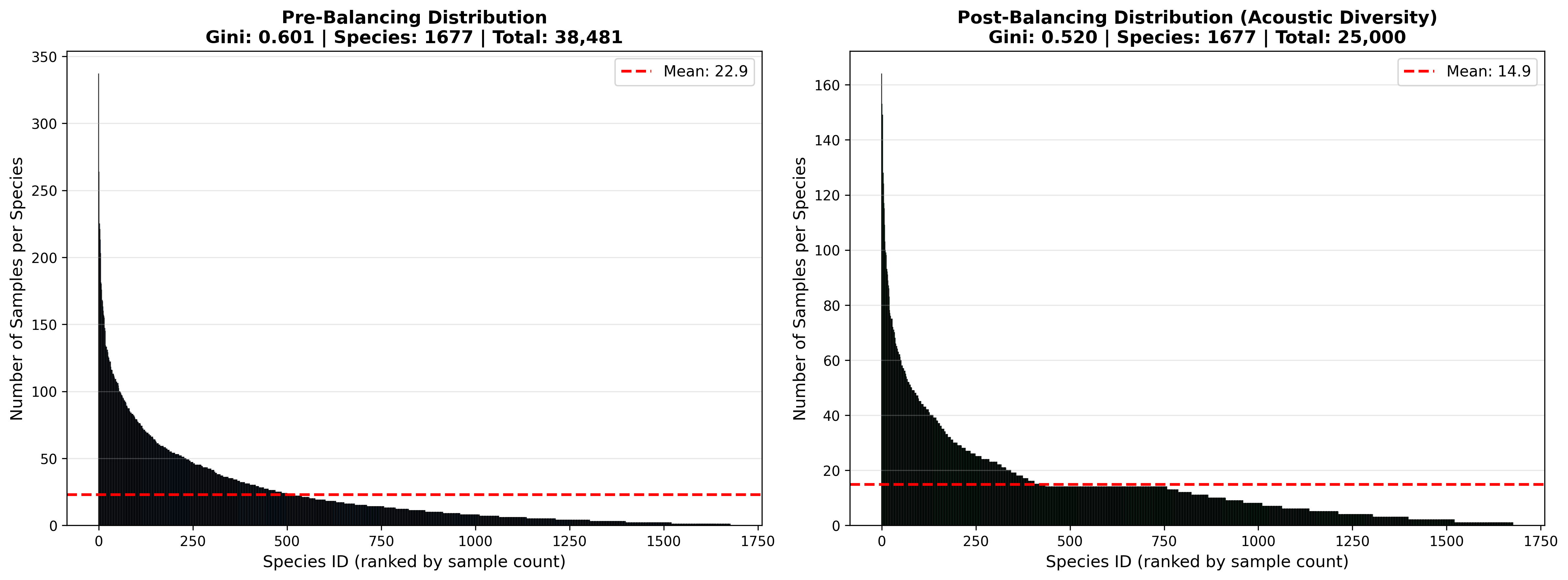}
  \caption{Species distribution before and after diversity-aware balancing.
    Pre-balancing (left): 38{,}481 clips across 1{,}677 species, Gini
    coefficient 0.601, mean 22.9 clips/species. Post-balancing (right): 25{,}000
    clips preserving all 1{,}677 species, Gini coefficient 0.519 (13.7\%
    reduction), mean 14.9 clips/species, yielding a more uniform distribution.}
  \label{fig:species_balance}
\end{figure*}

The deduplication stage identified 8 perfect duplicate pairs and 5
near-duplicate pairs, removing 13 files (0.03\% of the corpus) and yielding
38{,}481 unique recordings. Applying the diversity-aware balancing algorithm
reduced this to a curated subset of 25{,}000 positive samples while preserving
coverage of all 1{,}677 species (Figure~\ref{fig:species_balance}). The mean
number of samples per species increased to 14.9 and the Gini coefficient
decreased from 0.601 to 0.519, corresponding to a 13.7\% reduction in
inequality.

The final dataset contains 3{,}553 unique acoustic clusters. The negative
corpus contains exactly 25{,}000 clips drawn from six source datasets, as
summarized in Table~\ref{tab:negative_sources}. Table~\ref{tab:mybad_statistics}
lists the key properties of the complete SEABAD dataset.

\begin{figure}[t]
  \centering
  \includegraphics[width=\linewidth]{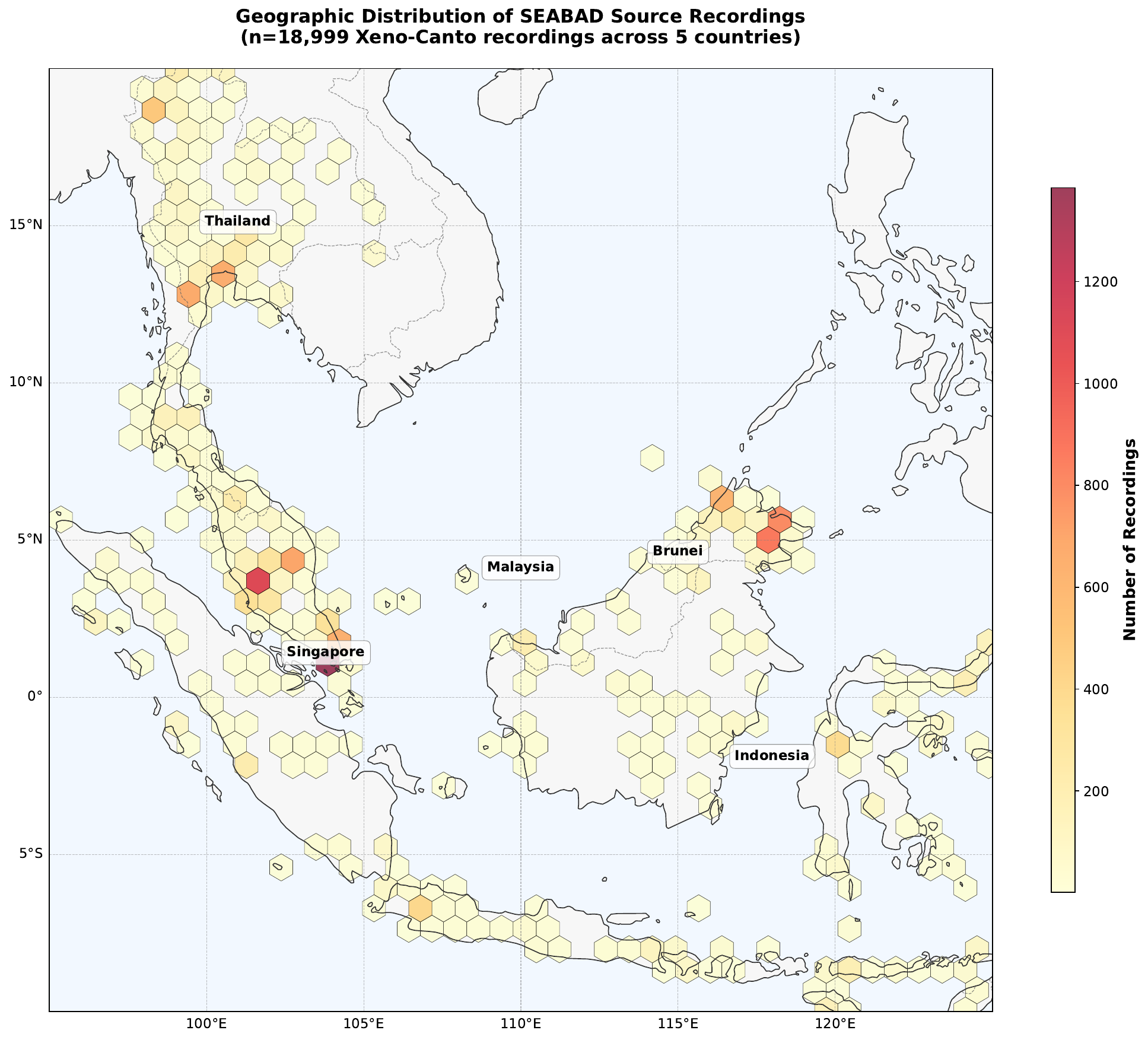}
  \caption{Geographic distribution of SEABAD positive recordings
    across Southeast Asia. Hexagonal bins show recording density (darker = more
    recordings). Shown are 18{,}999 clips within the defined SE Asia bounding box
    (latitudes $-9.97^\circ$ to $19.98^\circ$, longitudes $95.37^\circ$ to
    $124.88^\circ$), spanning five countries: Malaysia ($n=8{,}196$, 43.1\%),
    Thailand ($n=5{,}207$, 27.4\%), Indonesia ($n=4{,}188$, 22.0\%), Singapore
    ($n=1{,}376$, 7.2\%), and Brunei ($n=32$, 0.2\%). Of the remaining 6{,}001
    clips, 602 (2.4\%) lacked precise coordinates and 5{,}399 (21.6\%) had
    coordinates outside the bounding box. Recordings sourced from the Xeno-Canto
    community archive.}
  \label{fig:geo_distro}
\end{figure}

Geographic coverage of positive samples is concentrated in five countries
(Figure~\ref{fig:geo_distro}): of 18{,}999 clips falling within the defined
Southeast Asia bounding box (latitudes $-9.97^\circ$ to $19.98^\circ$,
longitudes $95.37^\circ$ to $124.88^\circ$), Malaysia accounts for 43.1\%,
Thailand 27.4\%, Indonesia 22.0\%, Singapore 7.2\%, and Brunei 0.2\%.
Of the remaining 6{,}001 clips, 602 (2.4\%) lacked precise coordinates and
5{,}399 (21.6\%) had coordinates outside the bounding box. The geographic
distribution before and after diversity-aware balancing is shown in
Table~\ref{tab:country_distribution}.

\begin{table}[t]
  \centering\footnotesize
  \caption{SEABAD dataset statistics and characteristics.}
  \label{tab:mybad_statistics}
  \begin{tabular}{@{} l r @{}}
    \toprule
    \textbf{Property} & \textbf{Value} \\
    \midrule
    Total clips                          & 50{,}000 \\
    Positive (bird presence)             & 25{,}000 \\
    Negative (bird absence)              & 25{,}000 \\
    Unique bird species                  & 1{,}677 \\
    Clip duration                        & 3 seconds \\
    Sample rate                          & 16\,kHz \\
    Bit depth                            & 16-bit \\
    Format                               & WAV (16-bit PCM) \\
    \midrule
    Mean samples per species             & 14.9 \\
    Gini coefficient (post-balancing)    & 0.519 \\
    Quality rating A/B                   & 92.1\% \\
    \bottomrule
  \end{tabular}
\end{table}

\begin{table}[h]
  \centering\small
  \caption{Geographic distribution of positive samples before and after
    diversity-aware balancing. Malaysia retains its proportional share
    (33.1\%$\to$33.6\%) whereas Thailand decreases (26.4\%$\to$24.0\%),
    reflecting prioritization of acoustic diversity and recording quality
    over uniform geographic sampling.}
  \label{tab:country_distribution}
  \begin{tabular}{lrrrr}
    \toprule
    \textbf{Country}
      & \multicolumn{2}{c}{\textbf{Pre-balancing}}
      & \multicolumn{2}{c}{\textbf{Post-balancing}} \\
    \cmidrule(lr){2-3} \cmidrule(lr){4-5}
      & \textbf{Samples} & \textbf{\%}
      & \textbf{Samples} & \textbf{\%} \\
    \midrule
    Indonesia   & 12{,}996 & 33.8\% & 9{,}155 & 36.6\% \\
    Malaysia    & 12{,}743 & 33.1\% & 8{,}400 & 33.6\% \\
    Thailand    & 10{,}169 & 26.4\% & 5{,}996 & 24.0\% \\
    Singapore   &  2{,}469 &  6.4\% & 1{,}388 &  5.6\% \\
    Brunei      &    104   &  0.3\% &    61   &  0.2\% \\
    \midrule
    Total       & 38{,}481 & 100\%  & 25{,}000 & 100\% \\
    \bottomrule
  \end{tabular}
\end{table}

\subsection{Quality assurance outcomes}

The first audit round (500 clips, random seed 42) identified 20 errors; the
second round (500 clips, random seed 43) identified 2 additional errors. Of
the 1{,}000 clips inspected, \textbf{978 (97.8\%)} were judged correct, while
\textbf{22 (2.2\%)} required remediation. The errors were categorized as:
wrong onset ($n=15$), noise dominated ($n=6$), and no bird ($n=1$). The most
common failure mode was incorrect onset detection, where RMS energy was
dominated by wind or rain bursts preceding the actual vocalization. The
combined error rate of 2.2\% (22/1{,}000) yields a final margin of error of
$e = 0.90$ percentage points at 95\% confidence, confirming dataset labeling
accuracy of \textbf{97.8\% ± 0.9\%}. The final dataset has 92.1\% of clips
rated as quality class A or B with improved mean salience (0.378 vs.\ 0.367
pre-balancing).

\subsection{Baseline validation performance}

Table~\ref{tab:baseline_validation} presents performance metrics for four
standard architectures and one zero-shot generalist on SEABAD's test set. All
four trained models achieved exceptional performance across three random seeds:
accuracy ranged from 99.49\% to 99.73\%, AUC from 0.9988 to 0.9996, with
seed-to-seed accuracy standard deviation below 0.25\% for all architectures.
The mean accuracy across all SEABAD-trained models and seeds was 99.60\%.

\begin{table*}[t]
  \centering\small
  \caption{Validation results on SEABAD test set ($n=5{,}000$).
    SEABAD-trained CNNs are averaged across three random seeds (42, 100, 786).
    BirdNET v2.4 is evaluated zero-shot at threshold $\tau=0.1$ (BirdNET
    default). The 30.95 percentage-point accuracy gap between BirdNET and
    MobileNetV3-Small quantifies the domain-shift cost of applying a
    temperate-trained generalist to Southeast Asian soundscapes.
    MobileNetV3-Small serves as the primary edge-deployment baseline.}
  \label{tab:baseline_validation}
  \begin{tabular}{lcccccc}
    \toprule
    \textbf{Model} & \textbf{Params}
      & \textbf{Accuracy} & \textbf{AUC}
      & \textbf{Precision} & \textbf{Recall} & \textbf{F1} \\
    \midrule
    MobileNetV3-Small$^{\dagger}$ & 1.1M
      & 99.57 ± 0.25\% & 0.9985 ± 0.0002
      & 0.9956 ± 0.0012 & 0.9957 ± 0.0008 & 0.9957 ± 0.0025 \\
    EfficientNetB0 & 4.4M
      & 99.49 ± 0.23\% & 0.9991 ± 0.0004
      & 0.9959 ± 0.0018 & 0.9939 ± 0.0051 & 0.9949 ± 0.0023 \\
    VGG16 & 14.9M
      & 99.61 ± 0.03\% & 0.9995 ± 0.0001
      & 0.9960 ± 0.0014 & 0.9963 ± 0.0010 & 0.9961 ± 0.0025 \\
    ResNet50 & 24.2M
      & 99.73 ± 0.02\% & 0.9992 ± 0.0003
      & 0.9965 ± 0.0013 & 0.9980 ± 0.0012 & 0.9973 ± 0.0019 \\
    \midrule
    BirdNET v2.4$^{\ddagger}$ & 6.5M
      & 68.62\% & 0.7819 & 0.6499 & 0.8072 & 0.7201 \\
    \midrule
    \multicolumn{7}{l}{$^{\dagger}$Primary baseline for edge deployment
      (intended use case).} \\
    \multicolumn{7}{l}{$^{\ddagger}$Zero-shot at threshold $\tau=0.1$
      (BirdNET default).} \\
    \bottomrule
  \end{tabular}
\end{table*}

MobileNetV3-Small, the primary edge-deployment baseline with 1.1M parameters,
achieved 99.57\,\% ± 0.25\,\% accuracy and AUC 0.9985\,±\,0.0002. Its
balanced metrics --- precision 0.9956\,±\,0.0012, recall 0.9957\,±\,0.0008 ---
indicate no systematic class bias. This represents a 22$\times$ parameter
reduction compared to ResNet50 (24.2M) and 13.5$\times$ compared to VGG16
(14.9M), with a performance gap of only 0.16\% accuracy relative to the best
model.

BirdNET v2.4 (\citealt{kahl2021birdnet}), a species classification model
trained on a global corpus covering roughly 6{,}000 bird species, was evaluated
zero-shot on the SEABAD test set. Predictions were converted to binary
bird-activity labels by thresholding the maximum per-clip species confidence
at BirdNET's default threshold $\tau=0.1$. BirdNET achieved 68.62\% accuracy
and AUC 0.7819, corresponding to a gap of 30.95 percentage points in accuracy
and 0.2166 in AUC relative to MobileNetV3-Small trained on SEABAD.

\section{Discussion}

\subsection{Dataset quality}

The consistently strong and stable performance across all four architectures
--- exceeding 99.4\% accuracy and 0.998 AUC across all models and seeds ---
indicates that SEABAD provides reliable training data with few annotation
inconsistencies. The binary detection task exhibits clear acoustic distinction:
bird calls produce tonal, harmonic structures with distinct temporal modulation
in mel-spectrograms, while the diverse negative class predominantly exhibits
broadband, non-tonal characteristics. This acoustic separability, combined
with the 97.8\% labeling accuracy confirmed by manual audit, confirms that
the automated curation pipeline produces high-quality training data without
requiring fully manual annotation at scale.

The acoustic salience score can serve as a lightweight proxy to flag likely
onset errors before manual review. Clips where RMS energy is dominated by
broadband transients (wind, rain bursts) rather than tonal content receive
systematically low salience scores. In practice, applying a salience threshold
(e.g., retaining clips in the top 80th percentile) substantially reduces the
proportion of onset errors in the candidate pool before auditing, lowering
manual review burden without sacrificing true-positive recall.

\subsection{Domain shift and generalist models}

The 30.95 percentage-point accuracy gap between BirdNET and MobileNetV3-Small
trained on SEABAD quantifies the domain-shift cost of applying a
temperate-trained generalist to Southeast Asian tropical soundscapes. BirdNET's
training corpus contains limited representation of acoustic conditions common
in tropical ecosystems, including dense insect biophony, overlapping
multi-species choruses, and region-specific vocalization patterns. These results
support the central motivation for SEABAD: robust bird activity detection in
tropical environments depends heavily on region-specific training data, not
only on model architecture or scale.

\subsection{Edge deployment context}

While MobileNetV3-Small achieves outstanding accuracy, its 1.1M parameters
and approximately 4.4\,MB memory footprint (FP32 weights) exceed AudioMoth's
256\,kB RAM constraint by 17$\times$. This quantifies the fundamental
challenge for microcontroller-based bioacoustic monitoring: achieving high
accuracy within extreme constraints ($<$100K parameters, $<$256\,kB RAM).
Initial work addressing this challenge directly on SEABAD (\citealt{zabidi2026seabadnet_code})
achieves AUC 0.9840 with only 979 parameters (6.8\,kB INT8) and 0.10\,ms
inference latency on ARM Cortex-M4 hardware, confirming that true
microcontroller-class deployment is feasible.

\subsection{Methodology transferability}

A central goal of this work is transferability. The dual-branch curation
pipeline --- positive-label stages covering acquisition, deduplication,
segment extraction, species balancing, and QA; and negative-label extractions
from six heterogeneous sources --- has been explicitly designed to generalize
beyond Southeast Asian avifauna.

\textit{Acoustic similarity-based deduplication.} By leveraging FAISS-based
nearest-neighbor search over mel spectrogram embeddings, redundant recordings
are systematically removed, mitigating train-test leakage.

\textit{Diversity-aware species balancing.} Stratified sampling combined with
acoustic clustering promotes representative coverage across species while
limiting over-representation from highly redundant recordings, directly
addressing the long-tail distribution characteristic of biodiversity data.

\textit{Multi-source negative curation.} The integration of environmental
sounds, curated datasets, and real-world soundscapes produces negative samples
that more closely resemble deployment conditions, enhancing model robustness
to both ecological and anthropogenic noise.

These methods are domain-agnostic and can be applied to other tropical regions,
temperate ecosystems, or non-avian taxa with minimal modification.

\subsection{Implications for tropical PAM}

SEABAD has direct implications for PAM in tropical environments. By providing
region-specific training data, SEABAD reduces the domain shift caused by
differences in species composition, acoustic structure, and background biophony
between temperate and tropical ecosystems. More broadly, SEABAD supports the
development of cascaded PAM systems in which a lightweight bird activity
detector acts as a front-end trigger for more computationally expensive
species-level classifiers. Such architectures can substantially reduce energy
consumption and processing overhead, extending the operational lifetime of ARUs
during long-term deployments.

The strong baseline performance ($>$99.6\% accuracy) should be interpreted in
the context of the dataset design. Positive clips were centered on high-energy
vocal segments using RMS-based extraction, while negative samples were drawn
from curated non-bird corpora. This reduces ambiguous boundary cases and
produces cleaner training examples than those typically encountered in
continuous field recordings. In practical deployments, models must still handle
more difficult conditions such as overlapping species, distant vocalizations,
environmental noise, and degraded recording quality.

\subsection{Limitations and future directions}

Although SEABAD addresses several gaps in tropical bird activity detection,
important limitations remain. The dataset is focused on Southeast Asian
ecosystems and may not generalize well to other tropical regions such as the
Amazon, Central Africa, or Oceania without additional adaptation. It also
relies heavily on community-contributed recordings from Xeno-Canto, which
introduce geographic and temporal biases: recordings are often concentrated
near accessible locations, biased toward active breeding periods, and more
likely to contain distinctive vocalizations.

The composition of the negative corpus is a further limitation. Current
negative samples are primarily derived from temperate or globally sourced
datasets, leaving some acoustic conditions common in Southeast Asian
environments underrepresented, particularly heavy monsoon rainfall, dense
cicada choruses, and primate vocalizations. These sounds may contribute to
false positives during real-world deployments. SEABAD is designed specifically
for binary bird presence--absence detection; species-level classification is
outside the scope of this dataset and is not an intended direction.

Several directions for future work remain open: expanding the negative corpus
with tropical-specific environmental sounds, conducting long-term field
evaluations under varying environmental conditions, systematic error analysis,
benchmarking against generalist systems such as Perch, and evaluating transfer
across geographic regions. Embedded deployment
validation on ARM Cortex-M hardware is addressed in companion work
(\citealt{zabidi2026seabadnet_code}).

\subsection{Conclusions}

SEABAD establishes the first large-scale bird activity detection dataset
specifically designed for Southeast Asian tropical ecosystems, comprising
50{,}000 automatically curated 3-second clips spanning over 1{,}600 species.
Beyond releasing this resource, we provide a complete, reproducible
dual-branch curation methodology addressing key challenges in biodiversity
dataset construction: acoustic deduplication to prevent train-test
contamination, diversity-aware species balancing to mitigate long-tail
distributions, and multi-source negative curation to ensure robust
generalization across deployment conditions.

Baseline validation with MobileNetV3-Small (99.57\% ± 0.25\% accuracy,
1.1M parameters) and three larger architectures ($>$99.4\% across all models
and seeds) demonstrates SEABAD's quality with minimal variance, confirming
that the binary detection task is highly learnable. Companion work
(\citealt{zabidi2026seabadnet_code}) further demonstrates that SEABAD supports
training of ultra-lightweight CNNs with under 1{,}000 parameters (6.8\,kB
INT8) achieving AUC 0.9840, validating the dataset's utility for true
microcontroller-class deployment.

By releasing SEABAD as an open-access resource alongside detailed curation
protocols, we aim to lower barriers for conservation practitioners and
accelerate development of autonomous, energy-efficient passive acoustic
monitoring systems in biodiversity hotspots worldwide.

\section*{Data availability}

The SEABAD dataset is publicly available on Zenodo (\citealt{zabidi2026seabad})
with complete documentation and metadata. The curation pipeline and
experimental code are available as open-source software
(\citealt{zabidi2026seabad_code}):

\begin{itemize}
  \item \textbf{Dataset}: \url{https://zenodo.org/records/18290494}
  \item \textbf{Curation scripts}: \url{https://github.com/mun3im/seabad}
\end{itemize}

The release includes 50{,}000 3-second WAV clips (16-bit PCM, 16\,kHz mono)
with train/validation/test partitions, XC recording ID mappings, preprocessing
and deduplication scripts, diversity-aware balancing algorithms, trained
baseline model weights, and complete provenance metadata including Xeno-Canto
IDs, coordinates, and licenses. All recordings are sourced from Xeno-Canto
under Creative Commons licenses (CC BY-SA, CC BY-NC-SA, CC BY-NC-ND, CC0).
Users of SEABAD must adhere to the respective Creative Commons licenses when
using or redistributing the dataset. The curation pipeline is implemented in
Python 3.10+ using librosa, numpy, scikit-learn, and FAISS.

\section*{Acknowledgments}

The authors thank the recordists who contributed field recordings to the
Xeno-Canto community archive. This research received no specific grant from
any funding agency in the public, commercial, or not-for-profit sectors.

\bibliographystyle{plainnat}
\bibliography{references}

\end{document}